# Field emission mechanism from nanotubes through gas ionization induced nanoscale surface charging

*Changhua Zhan[1], Zhongyu Hou*[1]*

[1]*National Key Laboratory of Micro/Nano Fabrication Technology, Key Laboratory for Thin Film and Microfabrication of Ministry of Education, Research Institute of Micro/Nano Science and Technology, Shanghai Jiao Tong University, Shanghai 200030, People's Republic of China;* [2]*Central Academe, Shanghai Electric Group Co., Ltd.*

**ABSTRACT**

Experimental evidences reported in this letter show that the gas ionization induced positive charge accumulation can lead to the electron field emission from carbon nanotubes (CNTs) in an electrode system with proper range of gap spacing, where the CNT film with ethocel was covered with ZnO nanorods. The hypothesis for illustration is suggested that: 1) the cosmic ray ionization frequency increases $10^8 \sim 10^{10}$ times due to the metastable population resulted from the interaction between the gases and the CNTs; 2) the flux of positive charges is enhanced in the converged field due to the ZnO nanostructures. The resulted positive charge local density is high enough to trigger the field emission of the CNTs. The methodology may be useful in particle detectors and ionization gas sensors.

*Corresponding author: zhyhou@sjtu.edu.cn

*Introduction:* The experiments discussed in this letter aim to introduce a methodology to correlate the irradiation ionization in gases with the surface charging induced field emission, which may lead to a measurement method for the ionizing source, perhaps also attractive to the particle detectors as a potential barrier modification way[1]. The charging of a non-metallic thin film could modify the electronic band structure at the electrode surface or subject the metal electrode beneath to an intensive electric field, resulting in electron emission at much lower external field than that of the Fowler-Nordheim tunneling ($10^9$V/m level)[2]. The charging could be resulted from different mechanisms, e.g., the electron bombardment in the Malter effect[3], the high energy irradiation in a particle detector[4], and the quantum tunneling through the barrier at the conductor-dielectric interface[5]. Potentially, the ionization of the gases in a metal-dielectric-gas structure could also result in the charging. However, in a gas discharge system, the dielectric barrier behavior is more competitive and always leads to the non-self-sustainable charging. An external ionization source is thus needed. As a natural source, the ionization frequency of the environmental cosmic rays in the lower altitudes ($v_c$) is about $1\sim3$ cm$^{-3}$s$^{-1}$[6]. This implies that the gradient across the dielectric layer covering a cathode, resulted from the positive charge accumulation due to the cosmic ray ionization must be vanishingly trivial. In a recent publication[7], it is suggested that the ionization yield of the environmental cosmic rays could be increased by $10^8\sim10^{10}$ orders of magnitudes in an electrode system with nanostructures (ESON) of proper gap spacing. Thus, the charges resulted from the cosmic ray ionization may lead to the field emission from the one-dimensional nanostructures (ONSs)[8], if the charge flux could be concentrated to further enhance the charge density upon the ONSs. In this letter, as shown in Fig.1a, an electrode configuration with carbon nanotubes (CNTs) and ZnO nanorods is introduced. In a CNT film (CNTF) cathode prepared by screen-printing method, part of the CNTs are expected to be exposed to the gases and the others covered by the residue ethocel matrix[9]. Thus, the former may act as the source of metastable population to enhance the cosmic ray ionization yield; and on the top of the latter, ZnO nanorods could be polarized and cause the positive charge concentration in the vicinity of the CNTs. We observed the phenomenological evidences and modeled the physical processes based on the argument developed as above-mentioned.

*Experimental*: As shown in the scanning electron micrograph of Fig.1b, the ONS cathode, prepared on a Cr/Pt covered glass plate and referenced as sample group A, consists of three layers: 1) CNTF, the bottom layer, prepared by screen-printing method[9]; 2) ethocel film, the intermediate layer, partially covering the CNTFs; 3) ZnO nanorods of ~200nm in average diameter and ~200nm in length, the surface layer, hydrothermally synthesized on the samples[10]. The other five controlled sample groups include: 1) CNTFs with ethocel insulating layer on Cr/Pt thin film of ~300nm in thickness (group B); 2) ~6μm long ZnO nanorods' array on Cr/Pt (group C); 3) ~200nm long ZnO nanorods on Cr/Pt (group D); 4) ~200nm long ZnO nanorods covering the ethocel layer on Cr/Pt (group E); 5) Cr/Pt thin film (group F). All the samples have been prepared on the quartz substrate and the opposite electrodes of the samples are the wet-etched silicon plates with array of ventilating holes of ~50μm in width, and the gap distance is kept at ~325μm by a polyimide film spacer. The discharge currents have been measured by Agilent 2911A in different volatile gas species mixed in the nitrogen as the variables, i.e., acetone, methanol, and ethanol. The current density measurement results have been normalized with respect to the applied electric field, i.e., the gaseous conductivity, say,

$G_{gs}$. The conductivity due to the polyimide spacer ($G_s$) has been considered in the measurements[7]. The samples have been finalized with atmospheric breakdown. The technical details of the sample preparation and the experimental setup may refer to the supporting information[11]. In the following, we shall extrapolate the field emission behavior in an ESON from the observed phenomenological pattern and illustrate the underlying mechanism based on the convergence band model[12].

*Results and discussions*: The *I-V* characteristic curve for the sample A is shown in Fig.2a, expect for the magnitudes, it is the same as the sample F where the first stage of quasi-linear growth and saturation is governed by cosmic ray ionization. The same mechanism is suggested for the quasi-linear stage of sample A. Given that the gaseous conductivity is determined by the charge flux in an electric field[6], it could be used as a measure of the gaseous ionization yield in different electrode systems. The gaseous conductivity of the group F ($G_{gs}^F$) is measured to be a constant in different gases[7] and assigned to the conductivity of the atmospheric air, which is ~$3.35 \times 10^{-15}$S/m measured by the Gerdien method[6]. Both of the conductivity of the sample group D and E behave similarly to the sample F. $G_{gs}$ of the sample group B ($G_{gs}^B$) and C ($G_{gs}^C$) are in the similar orders of magnitudes comparing to that of the group A ($G_{gs}^A$), but behave differently to the variation of the volatile species as shown in Fig. 2. According to the results shown in Fig. 2b, the gaseous conductivity in an ESON due to the cosmic ray ionization is at the level of $10^{-12}$S/m in the organic volatile mixtures with nitrogen.

According to the calculation of Laplace's equation using finite element method, the field enhancement effect of ONAs has resulted in the electric field whose maximum strength ($E_l$) is at the level of $10^6 \sim 10^7$V/m at 210V in the ethocel thin film, and of $10^5 \sim 10^6$V/m in the gases ranging less than 100nm. This weak-field-condition implies that the positive charges are not resulted from any self-sustainable processes of collisional ionization, field emission, or field ionization[13]. As shown in the schematics of Fig. 1a, we argue that *the surface charging induced field emission (SCFE) of the ONAs is possible when the positive charge density resulted from cosmic ray ionization in an ESON is high enough to enhance the local electric field strength*. Suppose that the positive charges have been totally collected in the corner region formed by the boundary between a ZnO nanoparticle and the ethocel film ($A_C$ region), where the field line converged, we could deduce the net positive charge accumulation equilibrium density, $\rho_V$, before the SCFE initiation, as follows.

In the $A_C$ region, the positive charge net flux density before the ignition of field emission, $i_p$, could be expressed as: $i_p = i_c - i_{diff} - i_{cond}$. It describes the source flux density, $i_c$, due to cosmic ray ionization considering the recombination processes, balanced by $i_{diff}$, the diffusion flux density, and $i_{cond}$, the conduction flux density through ethocel film. $i_c = (K_C V_a / \delta 2)\sqrt{4e^2 \mu_i \mu_e v_c S n_0 / \beta}$ [7,11] with the condition that the positive charge flux is consistent outside and inside the $A_C$ region, where $V_a$ is the applied voltage, $K_C$ is the dimensionless correction parameter, $S$ is the electrode area and $\delta S$ is the area of the $A_C$ region normal to the electrode surface, $e$ is the elementary charge factor, $\mu_i$ and $\mu_e$ are the drift mobility of positive ions and electrons[14], $n_0$ is the number density of the gases, $\beta$ is the recombination coefficient[14]. $i_{diff} = D_p \nabla \rho_V$ and $i_{cond} = G_{eth} \overline{E}_{AC}$, where $D_p$ is the diffusion coefficient with the assumption that $D_p$ and $\mu$ are constrained by the Einstein relation, $G_{eth}$ is the conductivity of the ethocel film[15], and $\overline{E}_{eth}$, numerically treatable, is the averaged field strength in the ethocel film. Suppose that the charge distribution is uniform across the electrode surface and the charge density is equal for both of the two polarity, we get: $i_e = -(\mu_e/\mu_i) \cdot i_p$, where $i_e$ is the electron flux density; then we get: $G_A^0 = (1 + \mu_e/\mu_i)(i_c - i_{diff} - i_{cond})/\overline{E}_{AC}$, where $\overline{E}_{AC}$ is the averaged field strength in the $A_C$ region and $G_A^0$ is the gaseous conductivity of sample A before SCFE initiation. Accordingly, without SCFE flux, we find:

$$G_A^0 = (1+\frac{\mu_e}{\mu_i})\frac{(K_C V_a/\delta 2)\sqrt{4e^2 \mu_i \mu_e v_c S n_0/\beta} - D_p \nabla \rho_V - G_{eth}\overline{E}_{eth}}{\overline{E}_{AC}}. \quad (1)$$

Supposed that $\rho_V = K_p z^\eta$, where $K_p$ is in C/m$^{3-\eta}$ and $z$ is the direction normal to the electrode surface, $\rho_V$ could be calculated based on the measurement of $G_A^0$ through the trial-and-error method. The theory implies that the SCFE behavior of the introduced ESON could be considered as an evidence of cosmic ray ionization yield increment due to the processes in an ESON. Based on the physical model derived as mentioned above, we can deduce the characteristic behavior of the sample group A and examine the phenomenological consistency as follows.

*1) Role of ZnO nanorods' convergence effect:* According to equation 1 and the FEM calculation, if there is no field convergence effect due to the ZnO surface layer as to the sample group B, $v_c$ should be in the order of $10^{20}$~$10^{22}$/m$^3$ to initiate the SCFE. This should have led to $G_B$ a thousand times higher than $G_A$[7], which is contrary to the observation shown in Fig. 2. Thus, under the same circumstances, if the SCFE is triggered in the sample A, the electron flux may result in a higher total current ($I_{cc}$) than that of the sample B. However, the

ZnO surface layer may weaken the interactions between the gases and the CNTs because the CNTF surface is partially blocked, i.e., $G_A^0 = \xi G_B$, where $\xi$ is the areal proportion of the ZnO nanorods on the CNTF surface, so that a lower $I_{cc}$ is expected when the SCFE is not triggered. Consequently, when SCFE is triggered, $I_{cc}$ of the sample A ($I_{cc}^A$) tends to be higher than that of the sample B ($I_{cc}^B$) and vice versa; this could be considered as an evidence of the initiation of the SCFE in sample A. As shown in Fig. 2, $G_{gs}^A > G_{gs}^B$ in all the organic volatile mixtures with nitrogen, which indicates that the temporally averaged $I_{cc}$ at the same $V_a$ satisfies the SCFE initiation condition therewith; whereas, $G_{gs}^A < G_{gs}^B$ in nitrogen and air. Then, why the condition is gas sensitive?

*2) Impact of the gas species:* Based on the physical model of equation 1, it is supposed that the reason why the SCFE initiation condition is gas sensitive is because the transport coefficients, $\beta$, and $v_c$ are gas sensitive[14], which can be referred to the supporting information[11]. If $v_c$ is determined based on the measurement results of $G_B$, it is given that $\rho_V$ is about $9.77 \times 10^4$, $9.23 \times 10^4$, $2.85 \times 10^5$, $2.01 \times 10^5$ and $1.75 \times 10^5 C/m^3$ for air, nitrogen, 4702.59 ppm methanol, 2137.88 ppm ethanol, and 9417.04 ppm acetone in nitrogen, respectively, through the calculations, when $\delta$=0.001, $\eta$=3, $K_p$=1.17C, $V_a$=210V, $\xi$=5%, and when $A_C$ region is the cubic box of 100nm in edge length. According to the calculation through finite element method (FEM) using the $\rho_V$ results, as shown in Fig. 3 for the methanol case, the field strength on the surface of the CNTF ($E_{nt}$) could be at the level of $10^9$V/m, so that the field emission from the CNTs could be possible. The thickness of the ethocel film, ranging in 5nm~1μm, determined by the scanning electron microscope (SEM) observations[9], could not be controlled in the experiments and was set to be 10nm herewith. In the calculations, five possible positional configurations of CNTs and the ZnO nanorods have been considered. The results for other gases are referred to the supporting imformation[11]. At the regions with the proper configurations, the SCFE electron flux may penetrate through the thin layer of ethocel into the gas and result in the increase of $I_{cc}$. Thus, the reason why the SCFE condition is gas sensitive could be explained by that the gas sensitive surface charging mechanism, modeled by equation 1, can lead to the differences in the surface charge level, modeled by $\rho_V$, which results in the differences in the SCFE initiation conditions and the net effect to the general electronic characteristics. Then, one may ask: which one plays the major role among those five gas sensitive parameters? Mathematically speaking, the major 'numbers' are $v_c$ in different gases which causes $i_p$ different, the trace organic volatiles generally cause little deviation from the gaseous electronic properties of the nitrogen[14]. It is suggested that the Jesse effect of cosmic ray ionization[16] underlies the increase of $v_c$ in the trace gas mixtures with nitrogen.

*3) Gas concentration and Jesse effect*: In the gas mixtures, the Jesse effect tends to increase the resulted charge flux[7]; this makes $v_c$ a concentration sensitive variable in a gas mixture. The instrumentation used in our experiments cannot realize pure organic volatile environments, the impact of the Jesse effect to the ionization yield is demonstrated by the $N_2/O_2$ and $N_2/Ar$ mixtures. As shown in Fig.2, the discharge current generally tends to be larger when the ONSs are positively biased; this could be explained by the metastable population is sourced in the vicinity of the ONSs and the density gradient should point outward from the ONS surface. As a result, the electric field that the electrons, generated by the interaction between cosmic rays and the metastable particles, must move through should be stronger in the positively biased cases. Thus, a higher drift current is expected because the electron drift coefficient averaged along the path is higher in stronger fields; for the electron attachment gases, the tendency is enhanced by the electron attachment coefficient is lower in stronger field[14]. Thus, if the ionization yields due to the Jesse effect is strong enough to trigger the SCFE process, it may lead to a higher discharge current than that of the positive case; this can be used as an identification of the initiation of SCFE. It is shown that the SCFE process in the concentration of 900ppm Ar in nitrogen may be identified in this way. However, as shown in Fig. 4b, $I_d$ of positively biased case is always larger, i.e., the SCFE in electron attachment gases may not be identified in this way, because the negative charge formation could significantly decrease the SCFE current portion in the measurements.

4) $I_{cc}$ *temporal developments*: As shown in Fig. 5, temporal development of $I_{cc}$ has been plotted for different gases. Generally, $I_{cc}$ tends to decrease with time; this is within expectation as a self-limiting behavior because the upper electrode is the micro-mesh silicon plate with the surface partially covered with the $SiO_2$ dielectric layer, as shown in Fig. 1a[12]. On the contrary, the $I_{cc}$-$t$ relationships for the controlled sample group D, E, and F are basically time-independent. This can be considered as another support to the hypothesis of increase in the gaseous ionization frequency. In the sample group A, the SCFE electron flux may alter the $I_{cc}$-$t$ relationship if the avalanche process is initiated in the $A_C$ region, as to the case shown in Fig. 5c.

In conclusion, an electrode system is introduced, where the functioning mechanisms are suggested that: 1) the metastable population is generated through the interactions between the

CNTs and the gases; 2) the ZnO nanorods on the surface enhance the charge flux convergence in the vicinity of the CNTs so that the electron field emission is possible under small biases; 3) the ionization mechanism is limited to the external irradiation, as to the cosmic rays recognized herewith based on the *I-V* characteristics and the analysis of the field distribution. An analytical model is developed for the analysis of the fundamental behaviors and illustration of the observations. The results support that the field emission in such a device could be correlated with the irradiation ionization characteristics, which may be used to develop a mechanism for irradiation detection. Besides, the experimental observations are consistent with the theoretical description of the species sensitive gaseous conductivity, which can be considered as a gas sensing methodology. It is noticed that the recognition of the cosmic rays as the only irradiation source is still lack of strong evidences, although the theories developed herewith are free from the specific irradiation source. For example, the excited states generated from the interactions between the ONSs and the gases may play important roles. The future works will be focused on these topics.

Acknowledgements: This work was supported by the Natural Science Foundation (Grant Nos. 60906053, 09ZR1415000, and 61274118) and Ministry Foundation (Grant Nos. 51308050309 and 9140A26010112JW0301).

FIG. 1 (Color online) (a) Device schematics; (b) SEM micrograph of the device surface.

FIG. 2 (Color online) (a) schematic for the *I-V* characteristics of all the samples; (b) multi-time averaged *I-V* curves of linear-fitting. (c) Gaseous conductivity of different devices in different gases, where $O_A$, $O_B$, $O_C$, $O_D$, $O_E$ are air, nitrogen, 4702.59 ppm methanol, 2137.88 ppm ethanol, and 9417.04 ppm acetone in nitrogen, respectively.

FIG. 3 (Color online) Finite element model of the field distribution of sample group A.

FIG. 4 (Color online) Averaged current in different gases of the sample group A.

FIG. 5 (Color online) Current-time measurement results in different gases.